# Analysis of DNA Chains by Means of Factorial Moments

Huijie Yang[1,+], Fangcui Zhao[1], Yizhong Zhuo[2], Xizhen Wu[2]

[1] Physics Institute, Hebei University of Technology, Tianjin 300130, China

[2] China Institute of Atomic Energy, Beijing 102413, P.O.X 275(18), China

**Abstract**

By means of the concept of factorial moments we examine DNA sequences from yeast to distinguish coding and non-coding regions. It is found that the factorial moments may be a powerful tool for analysis of DNA sequences.

## I. Introduction

DNA carries the genetic information of most living organisms. And the goal of genome projects is to uncover that genetic information. Hence, genomes of many different species, ranging from bacteria to complex vertebrates, are currently being sequenced. As automated sequencing techniques have started to produce a rapidly growing amount of raw DNA sequences, the extraction of information from these sequences becomes a scientific challenge. A large fraction of an organism's DNA is not used for encoding protein [1]. Hence, one basic task in the analysis of DNA sequences is the identification of coding regions. Since biochemical techniques alone are not sufficient for identifying all coding regions in every genome, computational tools based on concepts used in many science fields have recently played a prominent role. To cite a few examples, we could mention gene identification [2-3], assignment of tentative functions to particular sequences [4-7], and elucidation of their structure [7-18]. A relevant contribution is due to statistical methods, namely Markovian approximations [19], correlation functions, and Fourier transform [7,9,10], etc. However, these methods do not give specific information of how different regions are characterized, and also fail to distinguish one given species from another. For instance, Markovian approximations describe a genome in terms of *k*-tuple overlapping series of nucleotides (where *k* is the Markovian order) and might ignore some correlations. Fourier transforms only detect periodicity and possible correlations, but the

---

[+] Corresponding author    E-mail: huijieyangn@eyou.com

information associated with these correlations lacks relevant details about the composition of DNA chains. On the other hand, scientists in the field are trying combinations of different methods for the recognition of coding and non-coding DNA regions (based on techniques such as those mentioned above) in order to improve the accuracy for prediction of different packages, which actually reach approximately 90% of accuracy [20-21]. What is more, the large amounts of statistical patterns that are different in coding and non-coding DNA have been found to be species dependent [22]. That is to say, traditional coding measures based upon these patterns need to be trained on organism-specific data sets before they can be applied to identify coding DNA. This training set dependence limits the applicability of traditional measures, as new genomes are currently being sequenced for which training sets do not exist. For these reasons, alternative tools able to give different ideas and estimators concerning the structure of DNA chains, especially the statistical patterns that are species-independent, represent an important contribution in the field. Detailed investigations on the statistical behaviors of DNA, especially on the differences between coding and non-coding segments make it possible for us to find methods to identify coding segments from DNA sequences theoretically [22-30]. Several novel methods have been suggested in literature, such as entropy segmentation, NM method, mutual information function, etc. [22-24].

According to Li [31], a gene is a sequence of genomic DNA or RNA that performs a specific function, a vague definition comparing with the traditional one. Performing the function may not require the gene to be translated or even transcribed. Three types of genes are recognized at present, e.g., protein-coding genes, RNA-specifying genes and regulatory genes. In this present letter the coding segments refer to protein-coding genes.

In this letter, we suggest the concept of factorial moments as a coding measure. By means of the concept of factorial moments we try to identify coding and non-coding regions of DNA sequences from yeast. This method uses only the known statistical general properties of coding and non-coding segments of DNA. In this way, the prior training on known data sets is avoided; furthermore the search for additional biological information (such as splice sites or termination signals) can also be avoided.

## II. Factorial Moments (FM)

More than ten years have witnessed a remarkably intense experimental and theoretical activity in search of scale invariance and fractal in multihardron production processes, for short also called "intermittency" [32]. The primary motivation is the expectation that scale invariance or self-similarity, analogous to that often encountered in complex non-linear systems, might open new avenues ultimately leading to deeper insight into long-distance properties of QCD and the unsolved problem of colour confinement.

Generally, intermittency can be described with the concept of probability moment (PM). Dividing a region of phase space $\Delta$ into $M$ bins, the volume of one bin is then $\delta = 1/M$. And the definition of $q$-order PM can be written as [33],

$$C_q(\delta) = \sum_{m=1}^{M} p_m^q,$$

Where $p_m$ is the probability for a particle occurring in the $m$'th bin, which satisfies a constrained condition, $\sum_{m=1}^{M} p_m = 1$. For a self-similar structure, PM will obey a power law as,

$$\lim_{\delta \to 0} C_q(\delta) \propto \delta^{(q-1)D_q},$$

And $D_q$ is called $q$-order fractal dimension or Renyi dimension. Simple discussions show that $D_0, D_1$ and $\{D_q | q \geq 2\}$ reflect the geometry, information entropy and particle correlation dimensions, respectively.

It is well known that intermittency is related with strong dynamical fluctuations. But the measurements for multihardron production obtain the distribution of particle numbers directly instead of the probability distribution. And the finite number of cases will induce statistical fluctuations. To describe the strong dynamical fluctuations and dismiss the statistical fluctuations effectively, factorial moment (FM) is suggested to investigate intermittency [34,35]. The generally used form for FM can be written as,

$$F_q = M^{q-1} \sum_{m=1}^{M} \frac{<n_m(n_m-1)...(n_m-q+1)>}{<n>^q},$$

Where M is the number of the bins the considered interval being divided into, $n_m$ the number of particles occurring in the *m*'th bin, and *n* the total number of particles in all the bins. A measure quantity can then be introduced to indicate the dynamical fluctuations,

$$\phi_q = \lim_{\delta \to 0} \frac{\ln F_q}{\ln(1/\delta)}.$$

Here we present a simple argument for the ability of FM to dismiss statistical fluctuations [33].

The statistical fluctuations will obey Bernoulli and Poisson distributions for a system containing uncertain and certain number of total particles, respectively. For a system containing uncertain total particles, the distribution of particles in the bins can be expressed as,

$$Q(n_1, n_2, ... n_M | p_1, p_2, ... p_M) = \frac{n!}{n_1! n_2! ... n_M!} p_1^{n_1} p_2^{n_2} ... p_M^{n_M}.$$

And $(p_1, p_2, ... p_M)$ are the probabilities for a particle occurring in the $1, 2, ..., M$ bins, respectively. Hence,

$$\langle n_m(n_m - 1)...(n_m - q + 1) \rangle$$
$$= \int dp_1 dp_2 ... dp_M P(p_1, p_2, ... p_M) \times \sum_{n_1} ... \sum_{n_M} Q(n_1, n_2, ..., n_M | p_1, p_2, ... p_M)$$
$$\times n_m(n_m - 1)...(n_m - q + 1)$$
$$= n(n-1)...(n-q+1) \times \int dp_1 dp_2 ... dp_M P(p_1, p_2, ..., p_M) p_m^q$$
$$= n(n-1)...(n-q+1) \langle p_m^q \rangle.$$

That is to say,

$$F_q(M) = C_q(M) \propto M^\phi, |M \to \infty.$$

Therefore FM can describe the strong dynamical fluctuations and can dismiss the statistical fluctuations effectively.

Besides the statistical fluctuations, there are some trivial dynamical processes that need to be dismissed. These trivial dynamical processes induce the average numbers of particles in different phase space bins being not same, and the form of FM should be the original one, which reads,

$$F_q = M^{-1} \sum_{m=1}^{M} \frac{<n_m(n_m-1)...(n_m-q+1)>}{<n_m>^q},$$

A typical method to dismiss the fluctuations due to this kind of trivial dynamical processes is to transform the original distribution to homogeneous distribution by means of integrate method as follows [36],

$$x(y) = \frac{\int_{y_a}^{y} f(y)dy}{\int_{y_a}^{y_b} f(y)dy}.$$

But in this paper we resolve this problem by constructing a series of delay register vectors based upon the DNA sequences, as illustrated in the next paragraph.

### III. Application to DNA analysis

The concept of FM has been used to deal with many kinds of complex dynamical processes in physics, such as multi-particle production at high energy, DNA melting and denaturalization with the temperature increasing, etc. [37-38]. What is more, this concept is also improved to a new version called etermittency, to deal with some problems where statistical average can not be complemented properly [39].

Detailed works predict that in non-coding DNA sequences the elements A, T, C and G are not positioned randomly, but exhibit self-similar structure, while in coding DNA sequences the elements are distributed in a quasi-random way. Therefore, it may be a reasonable idea to distinguish coding and non-coding DNA sequences using the concept of FM.

There are several statistical features that can be employed to distinguish non-coding and coding regions, as illustrated below [40,41],

(a) The usage of strongly bonded nucleotide C-G pairs is usually less frequent than that of weakly bonded A-T pairs;

(b) The C-G concentration may differ significantly between organisms, but is generally larger in coding than in non-coding regions.

(c) The C-G concentration makes a strong "background" contribution to any possible differences between non-coding and coding subsequences.

(d) Non-coding regions display long-range power-law relations, and have common features to hierarchically structured languages, i.e. a linear Zipf plot and a non-zero redundancy. That is to say, there are deterministic structures in non-coding regions.

While for coding regions, it seems that random rules dominate the sequences.

Therefore, the coding and non-coding regions behave different completely. They are sequences obeying different laws. To take into account these statistical characteristics of DNA sequences, we construct a process as illustrated below [42-46],

(a) *d* successive nucleotides along a DNA sequence are regarded as a case containing d particles. The state of the case can be described with a d-dimensional vector as $(x_1, x_2, x_3...x_d)$, where $x_i$ is the state value for the $i'th$ nucleotide. We can define the state values according to our counting rules. In this paper $x_i$ is set to be *1* when the $i'th$ position is occupied with C or G, and *0* for A or T.

(b) For a segment with length *N*, the total possible $N-d+1$ successive cases form a process. The process covers the entire DNA segment we are interested in, which can be expressed with the series in d-dimensional delay-register vectors:

$$(x_1, x_2, x_3...x_d)$$

$$(x_2, x_3, x_4...x_{d+1})$$

$$\downarrow$$

$$(x_{N-d+1}, x_{N-d+2}, x_{N-d+3}...x_N)$$

For each case we can reckon the number of occurrences of the nucleotides C and G. Then the density spectrum $\rho_m$ (i.e., *m* distribution, normalized to unity) is obtained based upon the number of occurrences in all the cases. In this paper $F_q$ with q=4 are calculated. Obviously $F_q$ with other values of q, such as 5,6, can be gained easily if necessary. To indicate the differences of processes constructed above which reflect the behaviors of different regions in DNA sequence, we introduce a measure quantity as below,

$$\Delta F(t) = \sqrt{\frac{\sum_m (F_{0m} - F_{tm})^2}{\sum_m 1}},$$

Where *m* is the length of a case, $F_{0m}$ and $F_{tm}$ are FM of the initial process (i.e. the region for reference) and the $t'th$ process, respectively. If the $t'th$ region behaves similar with the initial one $\Delta F(t)$ will tend to zero, while $\Delta F(t)$ will be a definite non-zero value when the successive processes step into a region obeying different laws comparing with the initial one. What is more, two regions with similar behaviors will have almost same values of $\Delta F(t)$.

In Fig.(1) we shows the results for DNA sequences from Yeast. The unitary values of $\Delta F(t)$ are presented here. The initial part 1-1200bp is chosen to be the reference segment. The length of a case is set to be 10,20,30,40,50,60, respectively. The length of a segment used to construct a process is 1200bp. We can find that the right borders for almost all the coding regions occur at the bottoms of valleys. Because we can get the positions of valleys with a considerable precision, the right borders can be determined with the FM appropriately.

In Fig.(2) the left borders are determined. Firstly the considered DNA sequence is arranged in an inverse order, e.g., numbering the initial DNA sequence denoted with *1,2,3…N* with *N, N-1, …1*. Then the unitary $\Delta F(t)$ values are calculated. The positions of valleys can fit with the left borders very well.

Here we meet an essential problem, that is, how can we find a proper segment of DNA sequence to be employed as reference. Bad reference may induce fuzzy results. Investigations on the differences among coding segments or non-coding segments may be helpful, and the FM method is clearly a powerful tool. It is interesting to find in the results above that the right borders or the left borders are almost all positioned around two typical values, respectively. Perhaps we can catalogue the borders according to the quantity $\Delta F(t)$ in a certain degree.

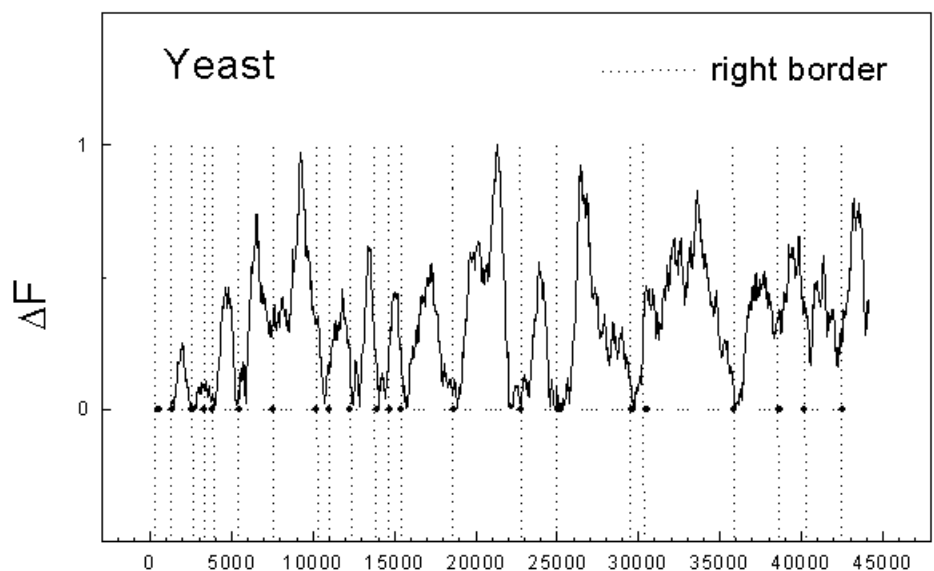

Right borders for DNA sequence from Yeast

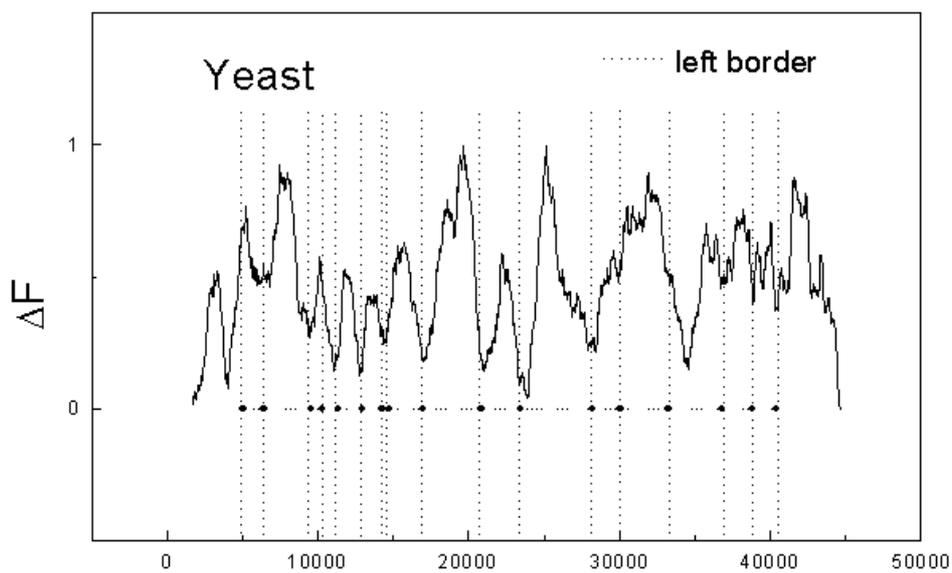

Left borders for DNA sequence from Yeast